\documentstyle{article}

\input epsf

\begin{document}

\title{DYNAMICAL GENERATION OF THE GAUGED SU(2) LINEAR SIGMA 
  MODEL\thanks{
   Published in {\it Mod.\ Phys.\ Lett.}\/ {\bf A10} (1995) 251.}}
\author{R.  DELBOURGO and M.  D.  SCADRON\thanks{ Permanent address:
    Physics Department, University of Arizona, Tucson, AZ 85721, USA} \\
    {\em Physics Department, University of Tasmania, Hobart,
    Australia 7005}} \date{15 October 1993}

 \maketitle
\begin{abstract}
The fermion and meson sectors of the quark-level SU(2) linear sigma model
are dynamically generated from a meson-quark Lagrangian, with the quark
($q$) and meson ($\sigma$, $\vec{\pi}$) fields all treated as elementary,
having neither bare masses nor expectation values. In the chiral limit,
the masses are predicted to be
$m_q =f_\pi g,~m_\pi=0,~m_\sigma=2m_q$, and we also find that the
quark-meson coupling is $g=2\pi/\sqrt{N_c}$, the three-meson coupling is
$g'=m_\sigma^2/2f_\pi=2gm_q$ and the four-meson coupling is
$\lambda=2g^2=g'/f_\pi$, where $f_\pi \simeq 90$ MeV is the pion decay
constant and $N_c=3$ is the colour number.
By gauging this model one can generate the couplings to the vector mesons
$\rho$ and $A_1$, including the quark-vector coupling constant
$g_\rho=2\pi$, $g_{\rho\pi\pi}$, $g_{A_1\rho\pi}$ and the masses
$m_{\rho}\sim 700$ MeV, $m_{A_1}\simeq \sqrt{3} m_{\rho}$; of course
the vector and axial currents remain conserved throughout.
\end{abstract}


\section{Introduction}

Although quantum chromodynamics (QCD) may indeed be the correct
chiral-invariant field theory of strong interactions, it is quite
difficult to quantify in the low-energy region. As an alternative,
one may consider effective chiral theories which presumably simulate
QCD in the infrared domain. Focussing on an isospin group, we will
specifically consider $0^{-+}$ pseudoscalars ($\vec{\pi}$), $0^{++}$
scalars ($\sigma$), $1^{--}$ vectors ($\vec{\rho}$) and $1^{++}$ axial
vectors ($\vec{A}_1$). Weinberg \cite{Wein} has recently reemphasized
the importance of treating all four of these chiral sectors in a global
context.

It is by now legend that the SU(2) linear sigma model (LSM) of Gell-Mann
and Levy \cite{GML}, with elementary nucleons, sigma and pion fields, is
the prototype field theory characterizing spontaneously broken chiral
symmetry.  Likewise the four-fermion (quark) field theory of Nambu and
Jona-Lasinio (NJL \cite{NJL}), generating the chiral-limiting (CL) bound
state $\bar{q}q$ meson masses,
\begin{equation}
m_\pi=0,\qquad m_\sigma=2m_q, \label{one}
\end{equation}
where $m_q$ is the quark mass, is the classic example of dynamically
broken chiral symmetry. These two field theories appear to be different
in as much as the nucleon level LSM does not constrain the scalar mass
$m_\sigma$ as does Eq.\ (\ref{one}), although the masslessness of the
Goldstone field $\vec{\pi}$ is certainly preserved in both cases.

However, if one formulates the SU(2) LSM {\em at the quark level}, it has
been suggested \cite{Egu} that with spontaneous breaking the theory may be
identical to the dynamically broken NJL model, for $N_c=3$. In fact, the
one-loop order LSM not only recovers (\ref{one}), but it also leads to
many more low-energy theorems \cite{HS,MDS}, known to hold in tree
order \cite{f2}. Notwithstanding the above successes, a spontaneously
broken LSM still appears fundamentally different from the dynamically
broken NJL theory, especially in its {\em effective description} of
low-energy QCD.

In this paper, in sections~\ref{sec:two} and~\ref{sec:three}, we will
dynamically generate the LSM at the quark level in the spirit of NJL.
Unlike others \cite{others} this will be done in the self-consistent
sense that {\em bare masses and expectation values vanish},
$$0 = m_{q0} = \langle\vec{\pi}_0\rangle = \langle\sigma_0\rangle, $$
signifying that the dynamically induced quantities $m_q$ and
$\langle \sigma \rangle = - f_\pi$ obey $\delta m_q=m_q,\delta f_\pi=f_\pi$.
We will likewise induce the cubic and quartic meson couplings by the
same non-perturbative bootstrap. Thus for our starting chiral quark
model (CQM) Lagrangian we take
\begin{equation}
 {\cal L} = \bar{\psi}[i\gamma.\partial + g(\sigma+i\vec{\tau}.\vec{\pi}
               \gamma_5)]\psi +
        [(\partial\sigma)^2 + (\partial\vec{\pi})^2]/2, \label{two}
\end{equation}
where $\sigma$ and $\pi$ are scalar and pseudoscalar meson fields and the
quark spinor $\psi$ corresponds to an up-down doublet coming in $N_c$
colours.

Following the approach of NJL, we induce a non-perturbative quark mass
$m_q$ by anticipating a mass gap equation $\delta m_q = m_q$ through a
quark loop, which also induces the pion decay constant. In this way the
associated axial vector current for the (now massive) free quarks,
\begin{equation}
 \vec{{\cal A}}_\mu = \bar{\psi}\gamma_\mu\gamma_5\vec{\tau}\psi/2
         + i f_\pi g(\partial_\mu/\partial^2)
             \bar{\psi}\gamma_5\vec{\tau}\psi, \label{three}
\end{equation}
develops a pseudoscalar pole and remains conserved in the chiral limit
(CL), $\partial^\mu\vec{{\cal A}}_\mu = 0$. Of course the Goldstone pion
\cite{GSW} remains massless and the Goldberger-Treiman (GT) relation must
hold at the quark level, $f_\pi g=m_q$ to ensure axial current conservation.
The CL non-perturbative pion decay constant $f_\pi\simeq 90$ MeV arises
from the definition
$\langle 0 |{\cal A}_\mu^j(0)|\pi^k(q) \rangle = i\delta^{jk}f_\pi q_\mu$
and $g$ is nothing but the dimensionless meson-quark coupling appearing
in (\ref{two}).

Then in section~\ref{sec:four} we gauge the chiral quark model (2) to
include vector and axial-vector couplings and masses in the theory.
Finally, in section~\ref{sec:five}, we give a dynamically generated
LSM interpretation of the very successful phenomenology of vector
meson dominance. The results are summarized in section~\ref{sec:six}.

\section{Generating the sigma model}
\label{sec:two}

In the NJL model, the mass gap and binding equations are equivalent but
both are quadratically divergent. However, in our quark-meson CQM scheme
(\ref{two}), the one-loop graph of Fig. 1a which induces
$f_\pi$ is logarithmically divergent, whereas the mass gap equation of
Fig. 1b is quadratically divergent \cite{f3}
In the former case we obtain in the soft limit \cite{f4}
$$ f_\pi= -4iN_cgm_q \int \bar{d}^4p/(p^2-m_q^2)^2.$$
Substituting the GT relation, the quark mass $m_q \neq 0$ cancels out,
leading to the logarithmically divergent equation,
\begin{equation}
1 = -4iN_cg^2\int \frac{\bar{d}^4p}{(p^2-m_q^2)^2} .\label{four}
\end{equation}
This ``gap equation'' (\ref{four}) is compatible with a cutoff approach,
as we will show later.
With the quadratically divergent mass gap equation, depicted by the
quark tadpole graph of Fig. 1b, the condition $\delta m_q=m_q$ (or
$m_0=0$) yields \cite{f4,f5},
\begin{equation}
 m_q = -\frac{8iN_cg^2}{m_\sigma^2}\int\frac{\bar{d}^4p\quad m_q}
                           {p^2-m_q^2}. \label{five}
\end{equation}
However it is dangerous to introduce an ultraviolet cut-off in a {\em
  quadratically} divergent graph since this depends sensitively on
shift of origin; in fact attempting to do so above produces an
imaginary $g$.  Moreover since $m_q$ does not appear in our initial
CQM lagrangian~(\ref{two}), one could shift the $\sigma$ field to
$\sigma \rightarrow \sigma - f_\pi$ so that~(\ref{two}) acquires a
chiral-broken kinetic part $\bar{\psi} (i \not\!\partial - f_\pi
g)\psi$ with $f_\pi g = m_g$.  If the latter mass is to be identified
with the quark tadpole of Fig~1b, it must be a counter-term mass of
opposite sign~\cite{f5} (relative to standard Feynman rules or to
mass renomalization involving a cutoff) in order that the original
$\sigma$ field is unshifted as in~(\ref{two}).

Instead we shall henceforth appeal to dimensionless
regularization~\cite{DL} and make no reference to a cutoff.  This
gives in $2l$-dimensions,
\begin{eqnarray}
\int \bar{d}^{2l}p/(p^2-m^2)^2 &=& i\Gamma(2-l)(m^2)^{l-2}/(4\pi)^l,
                         \nonumber  \\
\int \bar{d}^{2l}p/(p^2-m^2) &=& -i\Gamma(1-l)(m^2)^{l-1}/(4\pi)^l.
                         \nonumber
\end{eqnarray}
Therefore in the four-dimensional limit, one obtains the finite difference,
\begin{equation}
\int \bar{d}^4p\left[\frac{m^2}{(p^2-m^2)^2}-\frac{1}{p^2-m^2}\right]
  =\lim_{l\rightarrow 2}\frac{im^{2l-2}}{(4\pi)^2}[\Gamma(2-l)+\Gamma(1-l)]
  = -\frac{im^2}{(4\pi)^2}. \label{six}
\end{equation}
This lemma (\ref{six}) allows us to express the quadratic divergence
in (\ref{five}) in terms of a logarithmic one and gives
$$m_q = -\frac{8iN_cg^2}{m_\sigma^2} \int \frac{m_q\,\bar{d}^4p}{p^2-m_q^2}
   = -\frac{8im_qN_cg^2}{m_\sigma^2}\left[ \int \frac{m_q^2\,\bar{d}^4p}
       {(p^2-m_q^2)^2} + i\frac{m_q^2}{16\pi^2} \right], $$
\begin{equation}
 {\rm or} \qquad 1 =
ac{2m_q^2}{m_\sigma^2}\left[ 
                1 +\frac{g^2N_c}{4\pi^2}\right]. \label{seven}
\end{equation}
Note that the sign change of the gamma function sum in~(\ref{six}) in
turn cancels the minus sign of the counter-term quark mass
in~(\ref{five}).  In short the lemma (\ref{six}) renders the mass gap
equation (\ref{five}) compatible with (\ref{four}), since both terms
in (\ref{seven}) are positive definite.

Next we examine the meson self-energies to one-loop order, depicted in 
Figs. 2a, 2b and 3a, 3b. Given 
${\cal L}$ in (\ref{two}), the pion self-mass bubble 
and tadpole graphs are both quadratically divergent but their coefficients 
precisely cancel, as required by the Goldstone theorem \cite{HS,GSW},
\begin{equation}
 m_\pi^2 = 4iN_c[2
g^2 - 4gg'm_q/m_\sigma^2]\int \bar{d}^4p/(p^2-m_q^2)
       = 0, \label{eight}
\end{equation}
provided we dynamically require a $\sigma\pi\pi$ coupling
$g' = m_\sigma^2/2f_\pi$. This is just
what the Gell-Mann-Levy model stipulates, independently of the
$m_\sigma$ scale.

The analogous scalar meson self-energy graphs of Figs.\ 3a, 3b can
also be handled by dimensional regularization, in contrast to the
approach of refs. \cite{MDS}. Although $m_\sigma$ also does not appear
in the primary CQM lagrangian~(\ref{two}), the shifted nonzero
$\langle\sigma\rangle$ will induce an $m_\sigma$ via Figs~3 but with
$m_q \not = 0$ in the quark loops.  Again using a counter-term
$m_\sigma^2$ in the dimensionless regularization approach~\cite{f5},
the unshifted lagrangian~(\ref{two}) dynamically generates a scalar
mass $m_\sigma \not = 0$.  In particular, including the 3$\sigma$
Feynman combinatoric factor \cite{BWL} of 3!, the CL sum of the scalar
meson tadpole and bubble graphs of Figs. 3a, 3b gives~\cite{f6},
$$m_\sigma^2= 8iN_cg^2\int \bar{d}^4p\frac{(p^2+m_q^2)}{(p^2-m_q^2)^2} -
     48iN_c\frac{gg'm_q}{m_\sigma^2} \int\frac{\bar{d}^4p}{p^2-m_q^2}.$$
Using the finite difference (\ref{six}) and the same CL relation
$g'=m_\sigma^2/2f_\pi$, we find
\begin{equation}
 m_\sigma^2 = 16iN_cg^2\left[ \int \frac{m_q^2 \bar{d}^4p}{(p^2-m_q^2)^2}
         -\int \frac{\bar{d}^4p}{p^2-m_q^2}\right]
        = \frac{N_cg^2m_q^2}{\pi^2}. \label{nine}
\end{equation}
Solving the two identities (\ref{seven}) and (\ref{nine}), we discover
that the NJL mass $m_\sigma=2m_q$ has been dynamically generated and
that the meson-quark coupling constant is determined to be
$g=2\pi/\sqrt{N_c}$.

\section{Higher point functions to 1-loop order}
\label{sec:three}

Proceeding to the 3-point functions induced by the quark loop, the
CQM Lagrangian (\ref{two}) dynamically generates the graphs of Figs.
4a, 4b. In the zero momentum CL we obtain \cite{f6}
\begin{equation}
 g_{\sigma\pi\pi}=-8ig^3N_cm_q\int \frac{\bar{d}^4p}{(p^2-m_q^2)^2}
           = 2gm_q, \label{ten}
\end{equation}
upon using Eq. (\ref{four}). Similarly, in the chiral limit,
\begin{equation}
 g_{\sigma\sigma\sigma}=-8ig^3N_c\int \frac{3p^2m_q + m_q^3}
           {(p^2-m_q^2)^3} \bar{d}^4p = 6gm_q, \label{eleven}
\end{equation}
where we have just retained the dominant logarithmically divergent piece.
These results accord perfectly with the tree coupling relations (for
$m_\sigma = 2m_q$),
\begin{equation}
 2gm_q = m_\sigma^2/2f_\pi = g' \label{twelve}
\end{equation}
associated with the interaction $g'\sigma(\sigma^2+\vec{\pi}^2)$,
including the Feynman symmetry factors. The only point of note is that
the cubic couplings have been derived in a bootstrapped manner in as much
as ``loops reproduce trees''. These bootstrap concepts have a direct
affinity with renormalization conditions \cite{SW} $Z=0$ appropriate to
particles which are not elementary, but bound states of more basic fields.

Finally we generate the quartic meson-meson couplings dynamically \cite{f6}.
The Feynman graphs of Fig. 5 arising from (\ref{two}) lead to
logarithmically divergent integrals$^6$ for which we use equation
(\ref{four}) again:
\begin{eqnarray}
-2g^2&=&8iN_cg^4\int \frac{\bar{d}^4p}{(p^2-m_q^2)^2} \nonumber \\
          &=& g_{\sigma\sigma\sigma\sigma}/6 = g_{\pi\pi\pi\pi}/6
              = g_{\sigma\sigma\pi\pi}/2. \label{thirteen}
\end{eqnarray}
This can be interpreted perfectly as an interaction -$\lambda(\sigma^2 +
\vec{\pi}^2)^2/4$ where
\begin{equation}
 \lambda = 2g^2 = g'/f_\pi. \label{fourteen}
\end{equation}
Thus the quark loop of Fig.5 also bootstraps to a tree and
generates the quartic meson couplings dynamically.

Collecting our various results, we have established that with the
non-perturbative generation of the quark mass and pion decay constant,
connected by $m_q=f_\pi g$, there are induced dynamically the meson masses
$m_\pi=0,\quad m_\sigma=2m_q$ and the meson interactions
$$ g'\sigma(\sigma^2+\bar{\pi}^2) -
(\lambda/4)(\sigma^2+\bar{\pi}^2)^2 $$ for $\lambda = 2g^2$, together
with the original meson-quark interaction
$$ g\bar{\psi}(\sigma + i\gamma_5\bar{\tau}.\bar{\pi})\psi, $$
with $g'=m_\sigma^2/2f_\pi=2gm_q.$ This is identical in form with the
model of Gell-Mann-Levy, except that now the mass of the scalar meson is
dynamically fixed to the NJL value in (\ref{one}) and the strong coupling
for $N_c=3$ is also fixed to be
\begin{equation}
 g = 2\pi/\sqrt{N_c} = 3.6276, \label{fifteen}
\end{equation}
which is compatible with the ratio $m_q/f_\pi \simeq 3.6$, arising
from the GT relation. Alternatively, making use of the experimental
$g_{\pi NN} \simeq 13.4$ in $N=qqq$ and \cite{PDG} $g_A=1.2573$,
we may estimate $g = g_{\pi NN}/3g_A \simeq 3.55$. The mass values also
make good sense since the CL constituent quark mass is dynamically fixed
to be $m_q = 2\pi f_\pi/\sqrt{3} \simeq 325$ MeV, as might be expected
for a constituent quark. Lastly, the NJL $\sigma$ mass is
$m_\sigma = 2m_q \simeq 650$ MeV in the chiral limit, not incompatible
with observational signals \cite{Obs}.

Notwithstanding the impressive self-consistency of the scheme, we have
one final constraint to consider: because we have worked around the true
vacuum, we must verify that the chiral-symmetric vacuum expectation
values are satisfied. This means that we must check Lee's \cite{BWL} null
tadpole condition, taking into account the induced meson interactions.
Evaluating the graphs of Fig. 6, in the language of dimensional
regularization, we must verify that \cite{f4} the tadpole {\em sum}
vanishes:
\begin{equation}
\langle \sigma' \rangle = 0 =  -8iN_cgm_q \int \frac{\bar{d}^{2l}p}
    {p^2-m_q^2} + 3ig'\int \frac{\bar{d}^{2l}p}{p^2}
    +3ig' \int \frac{\bar{d}^{2l}p}{p^2-m_\sigma^2}.\label{sixteen}
\end{equation}
The first and third integrals in (\ref{sixteen}) scale respectively to
$m_q^2$ and $m_\sigma^2$ and of course we discard the massless tadpole
in dimensional regularization. Hence we find as $l \rightarrow 2$ that
a cancellation of the pole terms will occur providing
\begin{equation}
 N_c(2m_q)^4 = 3m_\sigma^4. \label{seventeen}
\end{equation}
Curiously, this seems to require that there be three colours, $N_c=3$
for two flavours. This conclusion parallels the chiral anomaly \cite{ABJ}
prediction of the $\pi^0 \rightarrow 2\gamma$ quark loop
amplitude in the chiral limit \cite{f7},
\begin{equation}
F_{\pi^0\gamma\gamma} = \alpha N_c/3\pi f_\pi, \label{eighteen}
\end{equation}
which leads to a decay rate (for 3 colours and 2 flavours) of
\begin{equation}
 \Gamma_{\pi^0\gamma\gamma} = m_\pi^3|F_{\pi^0\gamma\gamma}|^2/64\pi
                  \simeq 7.63\,\,{\rm eV}, \label{nineteen}
\end{equation}
quite close \cite{PDG} to the measured value of $7.74 \pm .55$ eV. Note
that using a quark loop, rather than a nucleon loop \cite{JS}, the
successful rate (\ref{nineteen}) can also be regarded as an LSM prediction.

Stated another way, in the CQM field theory, the NJL relation
$m_\sigma = 2m_q$ is initially generated via the quark mass gap of
Figure 1b. However the consequent $Z=0$ vertex condition helps to induce
dynamically the linear sigma model field theory, starting with (2).
At that level one obtains a vacuum expectation value for the full $\sigma$
field (not the bare $\sigma_0$) as $\langle\sigma\rangle = -f_\pi$.
One then shifts to $\sigma'$ as $\sigma=\sigma' + \langle\sigma\rangle
= \sigma' - f_\pi$. This corresponds to adding ``zero'' 3-point terms
to the Lagrangian (2) of the type $Z_{g'}\sigma'(\sigma'^2+\vec{\pi}^2)$
with compositeness condition $Z_{g'}=0$. One thereby obtains the
renormalized Lagrangian,
$${\cal L} = \bar{\psi}(i\gamma.\partial -m_q)\psi +
         g\bar{\psi}(\sigma'+i\gamma_5\vec{\tau}.\vec{\pi})\psi +
         g'\sigma'(\sigma'^2+\vec{\pi}^2) +
         (Z_{g'}-1)g'\sigma'(\sigma'^2 + \vec{\pi}^2) + .. $$
where the dots refer to ``zero'' 4-point Lagrangian terms. As usual one
determines the vertex renormalization constant $Z_{g'}$ by evaluating the
perturbative sum of 3-point plus quark loop graphs and thus recovers Eqns.
(10) and (11) upon setting $Z_{g'}=0$. Note that the above renormalized
CQM Lagrangian ``knows'' about the meson terms such as the trilinear ones.

If instead we dynamically induce the entire linear sigma model, then
the quark mass gap graph of Fig. 1b must be supplemented by $\sigma'$-
and $\vec{\pi}$- mediated quark self-energies and also $\sigma'$ and
$\pi$ tadpole graphs, which then all sum to zero. In this scenario the
NJL mass relation $m_\sigma = 2m_q$ is a consequence of the Lee
constraint $\langle\sigma'\rangle = 0$, or Eqns. (16) and (17). Then
because~(\ref{eighteen}) empirically fixes $N_c=3$, one sees
from~(\ref{seventeen}) that the NJL mass relation follows.

Thus far we have consistently adopted only one regularization scheme,
dimensional regularization. Nevertheless a cutoff version of (\ref{four})
would replace the right-hand side (rhs) by
$\ln(1 + \Lambda^2/m_q^2) -\Lambda^2(\Lambda^2+m_q^2)^{-1}.$
Setting the latter to unity in turn suggests $\Lambda/m_q \simeq 2.3$ or
$\Lambda \simeq 750$ MeV for $m_q = 2\pi f_\pi/\sqrt{3} \simeq 325$ MeV.
This cutoff sensibly separates the elementary quarks and mesons from the
known higher-mass $\bar{q}q$ SU(2) bound states at $\rho(770), \omega(783),
A_1(1260)$, etc.

A few concluding remarks are called for. It is also possible to construct
an effective LSM potential following the method of of Coleman and
Weinberg \cite{CW}but that scheme does not require the conditions

$m_\sigma = 2m_q$ or $g = 2\pi/\sqrt{3}$, although we should mention that
a combined LSM-NJL picture \cite{Holdom} has recently been linked to low-
energy QCD.

\section{Generating the vector masses and couplings}
\label{sec:four}

We now make the chiral quark model (CQM) Lagrangian (\ref{two})
invariant under {\em local}
SU(2)$\times$SU(2) by gauging the entire structure. In the standard way one
replaces ordinary derivatives by covariant derivatives,
\begin{equation}
 D\psi= [\partial + ig_\rho(\vec{V}-\vec{A}\gamma_5).\vec{\tau}
            + ie V^{em}] \psi, \label{One}
\end{equation}
\begin{equation}
 D\sigma = \partial \sigma - 2g_\rho\vec{A}.\vec{\pi},\quad
 D\vec{\pi} = \partial\vec{\pi} + 2g_\rho(\vec{\rho}\times\vec{\pi}
          + \vec{A}\sigma), \label{Two}
\end{equation}
and then writes down the (massless) gauge-invariant Lagrangian
\begin{equation}
 {\cal L} = \bar{\psi}[i\gamma.D
             + g(\sigma + i\gamma_5\vec{\tau}.\vec{\pi})]\psi
        + [(D\sigma)^2 + (D\vec{\pi})^2]/2
        - [\vec{F}_V^2 + \vec{F}_A^2]/4; \label{Three}
\end{equation}
\begin{equation}
 \vec{F}^V_{\mu\nu}=\partial_\mu\vec{\rho}_\nu-\partial_\nu\vec{\rho}_\mu
         +2g_\rho(\vec{\rho}_\mu\times\vec{\rho}_\nu
               + \vec{A}_\mu\times\vec{A}_\nu), \label{Four}
\end{equation}
\begin{equation}
 \vec{F}^A_{\mu\nu}=\partial_\mu\vec{A}_\nu - \partial_\nu\vec{A}_\mu
       +2g_\rho(\vec{\rho}_\mu\times\vec{A}_\nu +
            \vec{A}_\mu\times\vec{\rho}_\nu).  \label{Five}
\end{equation}
Even though the $\sigma$ meson and both $u$ and $d$ quarks acquire masses,
the vector and axial currents of course remain conserved, leading to
gauge-invariant kinematic structures in amplitudes. Nevertheless we will
show that the full propagators of the vector and axial gauge fields
($\rho$ and $A_1$) develop poles at finite masses through the
contributions of quark and meson loops.

But first we dynamically induce the gauge coupling constant $g_\rho$ in
(\ref{One},\ref{Two}). This comes about through the vector current matrix
element
\begin{equation}
 \langle0| {\cal V}^{em}_\mu |\rho^0(k) \rangle
   = e k^2\epsilon_\mu(k)/g_\rho ;\qquad k^2 = m_\rho^2, \label{Six}
\end{equation}
represented by the quark loop in Fig. 7 which evaluates to
\begin{equation}
 \langle0| {\cal V}^{em}_\mu |\rho^0_\nu(k) \rangle =
    -eg_\rho\Pi_{\mu\nu}(k^2,m_q^2) =
   eg_\rho(k^2g_{\mu\nu} - k_\mu k_\nu)\Pi(k^2,m_q^2), \label{Seven}
\end{equation}
with
\begin{equation}
 \Pi(k^2,m^2) = -8iN_c\int_0^1 d\alpha \int\bar{d}^4p\quad
    \alpha(1-\alpha)/[p^2 - m^2 + k^2\alpha(1-\alpha)]^2. \label{Eight}
\end{equation}
Proceeding to the soft (chiral) limit in the invariant amplitude
(\ref{Eight}), one is led to
\begin{equation}
 \frac{1}{g_\rho^2} = \Pi(0,m_q^2) = -\frac{8iN_c}{6} \int
      \frac{\bar{d}^4p}{(p^2-m_q^2)^2} = \frac{1}{3g^2}, \label{Nine}
\end{equation}
upon employing the gap equation (\ref{four}). However, we already know
that with three colours, $g=2\pi/\sqrt{3}$, so we deduce that in the chiral
limit,
\begin{equation}
 g_\rho = \sqrt{3} g = 2\pi. \label{Ten}
\end{equation}
This result has already been obtained by other methods \cite{CN}.

Although (\ref{Ten}) roughly approximates the data, one must admit that it
is 50\% too large since the observed $\rho \rightarrow e\bar{e}$ rate gives
$g_\rho^2/4\pi \simeq 2.01$. It indicates that we may improve on our
analysis by moving away from the soft chiral limit to the real $\rho$ mass
shell. By this subtraction procedure we arrive at a better approximation,
$$\frac{1}{g_\rho^2}-\frac{1}{4\pi^2}=\Pi(m_\rho^2,m_q^2)-\Pi(0,m_q^2)$$
\begin{equation}
  = -8iN_c\!\int_0^1\!\! \alpha(1-\alpha) d\alpha\! \int \!\!
    \left[\frac{\bar{d}^4p}{[p^2-m_q^2+m_\rho^2\alpha(1-\alpha)]^2} -
        \frac{\bar{d}^4p}{[p^2-m_q^2]^2} \right]. \label{Eleven}
\end{equation}
To be consistent with the gap equation (\ref{four}), cut-off at $\Lambda$
so that
\begin{equation}
 1 = \ln(1+\Lambda^2/m_q^2) - 1/(1+m_q^2/\Lambda^2), \label{Twelve}
\end{equation}
yielding $\Lambda \simeq 750$ MeV for $m_q \simeq 325$ MeV, we reexpress
(\ref{Twelve}) for $N_c=3$ as
$$\frac{1}{g_\rho^2}-\frac{1}{4\pi^2}=\frac{3}{2\pi^2}\!\!\int_0^1\!
                       \alpha(1-\alpha)d\alpha \!
   \left[\ln\!\!\left(1+\frac{\Lambda^2}{m_q^2-m_\rho^2\alpha(1-\alpha)}
       \right)-\ln\!\!\left(1+\frac{\Lambda^2}{m_q^2}\right)\right.$$
\begin{equation}
 \qquad\qquad\qquad\qquad\qquad\qquad\qquad \left.
 -\frac{\Lambda^2}{\Lambda^2+m_q^2-m_\rho^2 \alpha(1-\alpha)}+
  \frac{\Lambda^2}{\Lambda^2+m_q^2} \right]. \label{Thirteen}
\end{equation}
Assuming that $m_\rho \sim \Lambda \sim 750$ MeV, we can estimate the
integrals on the rhs of (\ref{Thirteen}) quite well by using the mean value
$\overline{\alpha(1-\alpha)} = 1/6$. In this way we find
$g_\rho^{-2} - (2\pi)^{-2} \simeq 0.46 (2\pi)^{-2}$
or $g_\rho^2/4\pi\simeq 2.15$, much closer to the experimental value.

The quark-meson coupling $g_\rho$ determines the $\rho\pi\pi$ and other
couplings through the gauge principle as Eqs. (\ref{Two}) and
(\ref{Three}) readily show:
\begin{equation}
 {\cal L} \supset
2g_\rho(\vec{\rho}^\mu\times\vec{\pi}.\partial_\mu\vec{\pi}
           + \sigma\vec{A}^\mu.\partial_\mu\vec{\pi}
           + g_\rho\sigma^2\vec{A}_\mu.\vec{A}^\mu
           + 2g_\rho\sigma\vec{\rho}_\mu\times\vec{\pi}.\vec{A}^\mu)
           + \ldots  \label{Fourteen}
\end{equation}
However it is very interesting and important to check that the meson
interactions are also consistently determined by the quark loops as in
Fig. 8. A simple calculation of those diagrams shows that in
the chiral limit \cite{HS}
\begin{equation}
 g_{\rho\pi\pi} = -4ig^2g_\rho N_c \int \bar{d}^4p/(p^2-m_q^2)^2 = g_\rho,
            \label{Fifteen}
\end{equation}
using the gap equation (\ref{four}). Of course (\ref{Fifteen}) also
conforms to vector meson dominance (VMD) universality \cite{JJS}, which is
approximately valid \cite{f8}.
The bootstrapping (\ref{Fifteen}) of the $g_{\rho\pi\pi}$ coupling can
also be interpreted as a vertex renormalization condition \cite{SW},
$Z_g = 0$, appropriate to particles which are purely bound states of more
basic fields; in this case we are presuming that $\rho$ is to be regarded as
a $q\bar{q}$ bound state \cite{f9}.

The vanishing of the wave-function renormalization constant $Z$ in the
$\pi-\gamma$ transition element is another way of understanding how
(\ref{Nine}) emerges. Thus the full inverse (off-diagonal) vector meson
propagator is
\begin{equation}
 {\Delta^{-1}}_{\mu\nu}^{(\rho\gamma)} = (-k^2g_{\mu\nu} + k_\mu k_\nu)
   [Z + g_\rho^2\Pi^{(\rho\gamma)}(k^2,m_q^2)], \label{Sixteen}
\end{equation}
and $Z$ is normally taken to cancel the infinite part of the self-energy:
\begin{equation}
 Z = 1 - g_\rho^2 \Pi^{(\rho\gamma)}(0,m_q^2). \label{Seventeen}
\end{equation}
Setting $Z=0$, we recover precisely (\ref{Nine}) in the chiral limit or
$g_\rho=2\pi$, again.

The other meson interactions arise from quark loops as they do through
the gauge principle (with identical zero vertex renormalization constants)
and thus it is rather easy to read off from (\ref{Fourteen}) what they will
be. In particular the transition amplitude $A_1 \rightarrow \rho\pi$ is
found by substituting the induced vacuum expectation value
$\langle\sigma\rangle = f_\pi$, producing the interaction
$4g_\rho^2f_\pi\vec{A}.\vec{\rho}\times \vec{\pi}$ and the decay rate
\begin{equation}
 \Gamma_{A_1^{+}\rho^{+}\pi^0}=p(4g_\rho^2f_\pi)^2/8\pi m_{A_1}^2\simeq 250
 {\rm ~MeV}, \label{Eighteen}
\end{equation}
for $m_{A_1} \simeq 1260$ MeV, $p \simeq 620$ MeV and $f_\pi \simeq
90$ MeV, not incompatible with experiment \cite{PDG}.

Next we turn to dynamical generation of the vector meson masses. Although
current conservation suggests that the $\rho$ mass may be zero (because
of the gauge invariant form (\ref{Sixteen}) of the two-point function and
the strong analogy with vacuum polarization in QED), we should not jump to
such a conclusion without examining the {\em full} self-energy and this
includes the contributions from $\pi$, $\rho$ and $A_1$ intermediate states.
The various contributions are depicted in Fig. 9, producing the
(diagonal) inverse $\rho$ propagator,
\begin{equation}
 {\Delta^{-1}}_{\mu\nu}^{(\rho\rho)} = (-k^2g_{\mu\nu} + k_\mu k_\nu)
      [Z + g_\rho^2(\Pi^{(q\bar{q})} + \Pi^{(\pi\pi)} + \Pi^{(A_1\pi)}
       + \Pi^{(\rho\rho)} + \Pi^{(A_1A_1)})]. \label{Nineteen}
\end{equation}
Since Higgs-like vector tranversality still holds for~(\ref{Nineteen})
even on the $\rho$ mass shell, the lhs of~(\ref{Nineteen}) must
vanish.  But to seek a zero on the rhs of~(\ref{Nineteen}) when $k^2 =
m^2_\rho$, we must also require the $Z=0$ compositeness
condition~\cite{SW}.  Collecting the various terms and making use of
(\ref{Seventeen}), we must check that
$$0 = 1 - 4ig_\rho^2 \!\int_0^1 \!d\alpha \!\int\!\bar{d}^4p\left[
     \frac{(1-2\alpha)^2}{[p^2+m_\rho^2\alpha(1-\alpha)]^2} +
     \frac{2(2g_\rho f_\pi)^2}{m_\rho^2[p^2-m_{A_1}^2\alpha
       +m_\rho^2\alpha(1-\alpha)]^2}
     \right. $$
\begin{equation}
  \qquad\qquad\qquad\qquad
     \left.-\frac{22/3}{[p^2-m_\rho^2+m_\rho^2\alpha(1-\alpha)]^2}
    -\frac{22/3}{[p^2-m_{A_1}^2+m_\rho^2\alpha(1-\alpha)]^2} \right].
    \label{Twenty}
\end{equation}
The unity corresponds to the quark contribution; the second term, also
positive, is that from a scalar massless particle \cite{f10};
the third term arises through the induced
$\rho A_1\pi$ coupling in (\ref{Fourteen}), while the fourth and fifth
terms are {\em negative} and associated with the purely vectorial
contributions, including the famous factor of $11C_2/3$ that plays such an
important role in asymptotic freedom \cite{AH}.

Because all of the above integrals are just logarithmically divergent and
scaled to the gap equation (\ref{four}) or (\ref{Twelve}), with an implied
cut-off $\Lambda \sim 750$ MeV, a consistent procedure, already used to
evaluate (\ref{Thirteen}), is to reexpress (\ref{Twenty}) as
$$0 = 1 + \frac{g_\rho^2}{4\pi^2}\int_0^1\!d\alpha \left[ (1-2\alpha)^2
      \left(\ln(1-\frac{\Lambda^2}{m_\rho^2\alpha(1-\alpha)}) -
       \frac{\Lambda^2}{\Lambda^2-m_\rho^2\alpha(1-\alpha)}\right)
       \right.$$
$$ \qquad\qquad \left. +\frac{8f_\pi^2g_\rho^2}{m_\rho^2}
   \left(\ln(1+\frac{\Lambda^2}{m_{A_1}^2\alpha-m_\rho^2\alpha(1-\alpha)})-
  \frac{\Lambda^2}{\Lambda^2+m_{A_1}^2\alpha-m_\rho^2\alpha(1-\alpha)}\right)
       \right. $$
$$\qquad\qquad\left.-\frac{22}{3}\left(\ln(1+\frac{\Lambda^2}{m_\rho^2
  (1-\alpha(1-\alpha))})-\frac{\Lambda^2}{\Lambda^2+m_\rho^2
  (1-\alpha(1-\alpha))}\right)\right. $$
\begin{equation}
 \qquad\qquad\left.-
 \frac{22}{3}\left(\ln(1+\frac{\Lambda^2}{m_{A_1}^2-m_\rho^2
 \alpha(1-\alpha)})-\frac{\Lambda^2}{\Lambda^2+m_{A_1}^2-m_\rho^2
  \alpha(1-\alpha)}\right) \right] .  \label{Twentyone}
\end{equation}
Taking the masses to be physical, with $\Lambda\simeq 750$ MeV, we may
evaluate the integrals numerically as a function of $\Lambda/m_\rho$
and search for a zero; in this way we estimate the $\rho$ mass to be
about 650 MeV. The interpretation of these computations is that the
vector mesons do indeed develop masses dynamically of the right
magnitude.  Note that the dynamical consequence of generating a rho
mass via Eqns.~(\ref{Nineteen})-(\ref{Twentyone}) in one-loop order is
already simulated by the KSRF relation~\cite{KSRF}.  The latter holds
in tree-order for our dynamically induced LSM which also generates the
VMD relations via $Z=0$ conditions.

Rather than repeat this (somewhat tedious) process for the axial $A_1$
meson, we observe from the third term of (\ref{Fourteen}) that for a
non-zero $m_\rho$,
\begin{equation}
 m_{A_1}^2 = m_\rho^2 + (2g_\rho f_\pi)^2. \label{Twentytwo}
\end{equation}
Invoking the numerically accurate KSRF relation \cite{KSRF}, $m_\rho^2
\simeq 2g_\rho^2 f_\pi^2$, we may conclude that
\begin{equation}
 m_{A_1} \simeq \sqrt{3} m_\rho \simeq 1300 {\rm ~MeV},\label{Twentythree}
\end{equation}
close to experiment but some way from the prediction $m_{A_1} =
\sqrt{2}m_\rho$ that comes from spectral-function sum rules \cite{Wei}.

\section{Vector meson dominance revisited}
\label{sec:five}

Having completed the dynamical generation of the gauged chiral quark
model, we will end by commenting on the (extremely successful) VMD
picture. Not only is VMD universality a consequence of the bootstrap
mechanism, but the tree structure of all VMD graphs then becomes
manifest since quark loops bootstrap to trees.
Furthermore $g_{\rho\pi\pi} \simeq g_\rho$ is roughly
satisfied by the theoretically derived value $g_\rho \simeq 2\pi$ in the
chiral limit.

Concerning the $\rho-\gamma$ analogy of VDM, the relation
between the two is governed by the quark loop, aside from coupling
constants; in particular an examination of the quark loops for the
processes $\rho \rightarrow \pi\gamma$ and $\pi^0 \rightarrow
\gamma\gamma$ shows that the associated amplitudes \cite{f11} determine the
empirical $g_\rho/e$ coupling ratio \cite{GMSW}:
\begin{equation}
 g_\rho/e = |2F_{\rho\pi\gamma}/F_{\pi^0\gamma\gamma}| \simeq 17.8
        \label{Twentyfour}
\end{equation}
using the measured rates$^9$ $\Gamma_{\rho\pi\gamma} \simeq 67$ keV,
$\Gamma_{\pi^0\gamma\gamma} \simeq 7.7$ eV. Again this ratio is close to
the measured $g_{\rho e \bar{e}}$ value and fine structure constant:
\begin{equation}
 g_\rho/e \simeq 5.03/0.3028 \simeq 16.6  \label{Twentyfive}
\end{equation}
Likewise the vertices $A_1\rho\pi$ and $A_1\gamma\pi$ are VMD related and
both contain a common gauge-invariant kinematic factor
$(k.k'g_{\mu\nu} - k_\mu k_\nu ')$. In this case \cite{BS} one finds
\begin{equation}
 g_\rho/e = |F_{A_1\rho\pi}/F_{A_1\gamma\pi}| \sim 15, \label{Twentysix}
\end{equation}
using the observed rates $\Gamma_{A_1^+\rho^+\pi^0} \sim 250$ MeV and
$\Gamma_{A_1\pi\gamma} \simeq 640$ keV. The approximate agreement between
the three ratios (\ref{Twentyfour}) - (\ref{Twentyfive}) emphasizes the
(quark loop) $\rho-\gamma$ analogies of VMD.

Finally we compare the measured \cite{Ame} pion charge radius $r_\pi
= (0.66 \pm 0.02)$ fm with the gauged CQM and VMD predictions. The former
CQM quarkloop is known to give \cite{TG}
\begin{equation}
 r_\pi^2 = N_c/(2\pi f_\pi)^2 = 1/m_q^2 \simeq (0.61 {\rm fm})^2,
       \label{Twentyseven}
\end{equation}
in the chiral limit, while the latter VMD value is
\begin{equation}
 r_\pi^2 = 6/m_\rho^2 \simeq (0.63 {\rm fm})^2. \label{Twentyeight}
\end{equation}
Suffice it to say that the measured radius lends further support to the
gauged CQM and the compatible VMD picture.

In fact the entire scheme
presented here, with dynamically generated (chiral limit) couplings
$g = 2\pi/\sqrt{3}, g_\rho = 2\pi$ and masses $m_q = 2\pi f_\pi/\sqrt{3}
\simeq 325$ MeV, $m_\rho \sim 750$ MeV and $m_{A_1} \simeq \sqrt{3}
m_\rho$, along with the induced $\rho\pi\pi$ and $A_1\rho\pi$ vertices,
are all consistent with the data.

\section{Summary}
\label{sec:six}

In this paper we have started from the simple chiral quark model
(\ref{two}) where all bare parameters vanish and dynamically generated
the entire sigma model Lagrangian in Sections~\ref{sec:two}
and~\ref{sec:three} by working to one-loop order.  As a bonus we
obtained we obtained the meson-quark coupling $g=2\pi/\sqrt{3}$ (a
sizeable value) and sensible quark and sigma mass values.  Then in
Section~\ref{sec:four} we gauged the scheme (again with zero bare
vector masses) to include the vector meson interactions. Here we found
the chiral-limiting quark coupling to be $g_\rho=2\pi$ and got
sensible masses for the vectors $\rho$ and $A_1$, as well as sensible
interactions to the other mesons. Lastly, in section~\ref{sec:five},
we showed that the gauged CQM recovers the entire structure of
(tree-level) VMD. The importance of bootstrapping loop graphs into
tree graphs for dynamically generated field theories cannot be
overstressed. This arises from $Z=0$ conditions for fields such as
$\sigma$ and $\rho$ which formally appear as elementary but are in
fact bound states.

In hindsight we see that quadratically divergent tadpole graphs,
(determined by dimensional regularization) are vital for studying the
dynamical generation of the LSM field theory. Even though the sum over all
tadpole graphs is zero, the individual contributions do not vanish by
themselves; our procedure ensures that the Goldstone pion mass vanishes by
guaranteeing that the quark, pion and $\sigma$ bubble plus tadpole
sums are zero {\em separately} \cite{HS}.
The whole structure is tightly-knit and self-consistent.

\section{ACKNOWLEDGEMENTS}

RD acknowledges partial support from the Australian Research Council. MDS
appreciates the hospitality of the University of Tasmania and the partial
support of the U.S. Department of Energy. Both authors thank R. Tarrach
for enlightening comments and A. Bramon for prior discussions.


\begin{description}
\item{Fig.1a} Logarithmically divergent quark loop for $f_\pi$.
\item{Fig.1b} Quadratically divergent graph for $m_q$.
\item{Fig.2a} Quark loop contribution to the pion self-energy.
\item{Fig.2b} Quark tadpole contribution to the pion self-energy.
\item{Fig.3a} Quark loop contribution to the sigma self-energy.
\item{Fig.3b} Quark tadpole contribution to the sigma self-energy.
\item{Fig.4a} Quark loop contribution to the $\sigma\pi\pi$ vertex.
\item{Fig.4b} Quark loop contribution to the $\sigma\sigma\sigma$ vertex.
\item{Fig.5} Quark loop contribution to the quartic meson vertices.
\item{Fig.6} The total tadpole contribution to $\langle\sigma'\rangle$.
\item{Fig.7} Quark loop contribution to the $\rho-\gamma$ transition
amplitude.
\item{Fig.8} Quark loop contributions to the $\rho\pi^+\pi^-$ vertex.
\item{Fig.9} Quark, pion, $\rho$ and $A_1$ contributions to the $\rho$ meson
      self-energy.
\end{description}

\setlength{\unitlength}{1cm}
\newpage

\begin{figure}
\begin{picture}(14,18)
\put (0.0,0.0){\epsfxsize=12cm \epsfbox{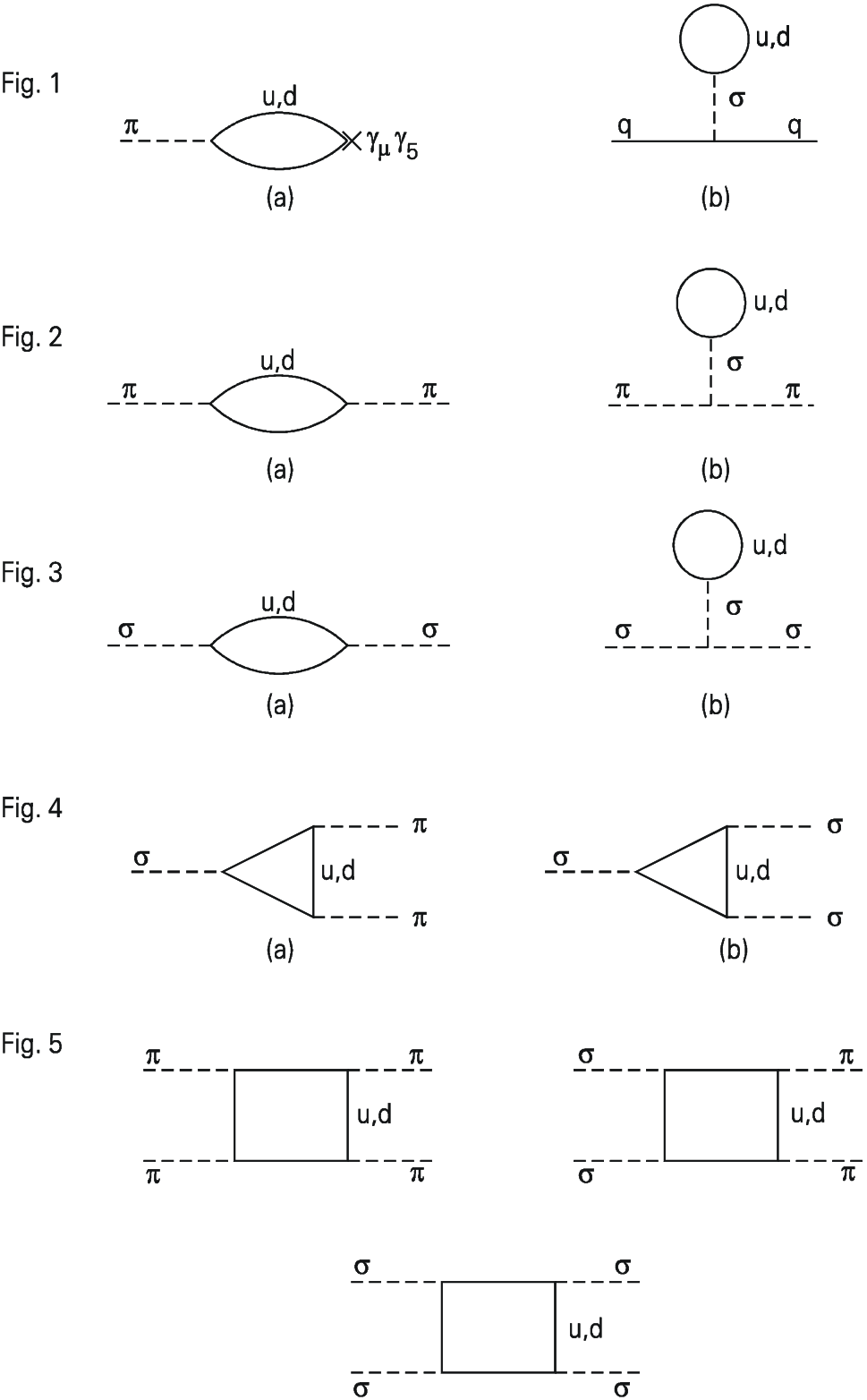}}
\end{picture}
\end{figure}

\newpage

\begin{figure}
\begin{picture}(14,18)
\put (0.0,0.0){\epsfxsize=12cm \epsfbox{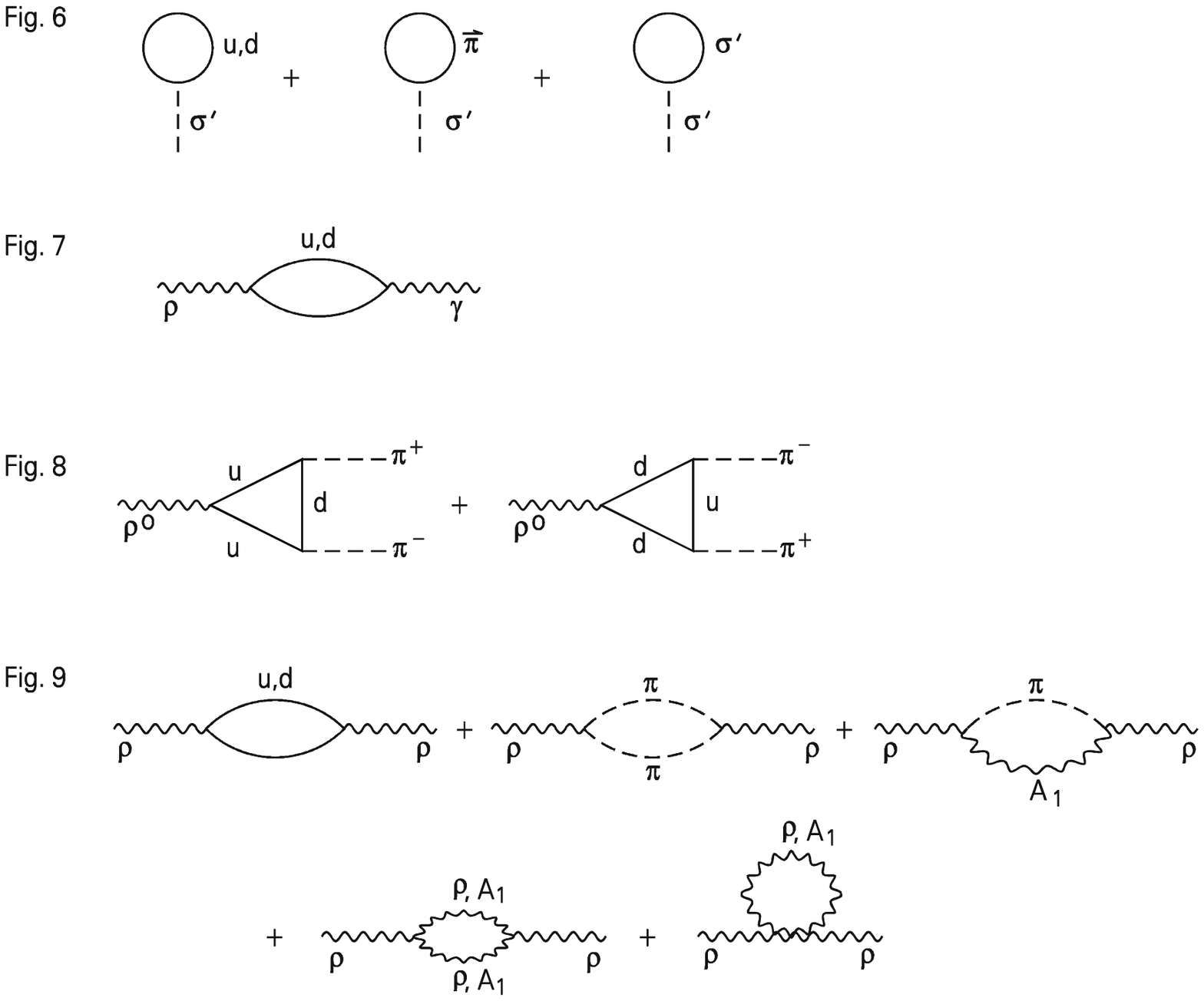}}
\end{picture}
\end{figure}

\begin{thebibliography}{99}
\bibitem{f3} We need not include logarithmically divergent ($\sigma$
  and $\vec{\pi}$ mediated) quark self-energy graphs in Fig.1b because
  direct calculation shows that they actually cancel on the quark mass
  shell due to chiral symmetry. However we must not consider
  quadratically divergent $\sigma$ and $\pi$ loop tadpole graph
  contributions because pure meson $\pi\pi\sigma$ and
  $\sigma\sigma\sigma$ couplings do not appear in the primary
  Lagrangian (\ref{two}). We will consider these additional tadpoles
  after dynamical generation of the LSM.
\bibitem{f4} Our notation is $\bar{d}^4p=d^4p/(2\pi)^4$ and we
 naturally just assume two flavours throughout.
\bibitem{f5} The emergence of an expectation value $f_\pi$,
 via $\delta f_\pi = f_\pi$ is virtually the same equation. Basically
 there occurs a cancellation between the physical quantity and the quantum
 correction, as happens for the fermion mass; the former is positive and the
 latter is negative. As such, eqs.(5) and (9) arise as counter-term
 masses.
\bibitem{f6} These integrals are dominated by their divergences
 and are thus being evaluated at zero external momenta.
\bibitem{f8} because $g_{\rho\pi\pi}^2/4\pi \simeq 2.9$, given the observed
 decay rate $\Gamma_\rho \simeq 150$ MeV.
\bibitem{f9} We invoked the same $Z_g=0$ condition when we boostrapped
 the $g'=g_{\sigma\pi\pi}$ coupling in the linear sigma model via a
 quark loop.
\bibitem{f10} That is why we have disregarded the pion tadpole
 bubble---zero by dimensional regularization.
\bibitem{f11} multiplying the kinematic factor
 $\epsilon_{\mu\nu\rho\sigma}k'^\mu k^\nu\epsilon'^\rho\epsilon^\sigma$,
 which is automatically gauge-invariant.
\end{thebibliography}
\end{document}